\documentclass[12pt,epsf,graphicx,epsfig]{elsarticle}
\usepackage{epsf}
\usepackage{epsfig}
\usepackage{graphicx} 
\usepackage{amsmath}
\usepackage{amsfonts,amsbsy}
\usepackage{amssymb}
\usepackage{showlabels}
\usepackage{graphicx}
\usepackage{bm}
\usepackage{resizegather}
\usepackage[colorlinks=true,pdfstartview=FitV,linkcolor=red,citecolor=blue,urlcolor=blue]{hyperref}

\def\be{\begin{equation}}
\def\ee{\end{equation}}
\def\bea{\begin{eqnarray}}
\def\eea{\end{eqnarray}}

\mathchardef\Re="023C
\mathchardef\Im="023D

\begin{document}

\title{\bf The Large N Limit with Vanishing Leading Order Condensate for Zero Pion Mass}

\author[int,ccnu]{Larry McLerran} 
\author[rbrc]{Vladimir Skokov}
\address[int]{Institute for Nuclear Theory, University of Washington, Box 351550, Seattle, WA, 98195, USA}
\address[ccnu]{China Central Normal University, Wuhan, China}
\address[rbrc]{RIKEN BNL Center, Upton NY}

\begin{abstract}
It is conventionally assumed that the negative mass squared term in the linear
sigma model version of the pion Lagrangian
is  $M^2 \sim \Lambda_{\rm QCD}^2$ in powers of $N_c$~\cite{Witten:1979kh}.  We
consider the case where $M^2 \sim \Lambda^2_{\rm QCD}/N_c$ so that to leading
order in $N_c$ this symmetry breaking term vanishes.  We present some arguments
why this might be  plausible.  One might think that such a radical assumption
would contradict lattice Monte Carlo data on QCD as  function of $N_c$.  We
show that  the linear sigma model gives a fair description of the data of
DeGrand and Liu  both for $N_c = 3$, and  for variable $N_c$.  The values of
quark masses considered by DeGrand and Liu, and by Bali et. al. turn out to be
too  large to  resolve the case we consider from that of the conventional large
$N_c$ limit~\cite{DeGrand:2016pur,Bali:2013kia}.  We argue that for quark
masses $m_{q} \ll  \Lambda_{\rm QCD}/N_c^{3/2}$, both the baryon mass  and
nucleon size scale as $\sqrt{N_c}$.   For $m_{q} \gg \Lambda_{\rm
QCD}/N_c^{3/2}$ the conventional large-$N_c$ counting holds.
The physical values of quark masses for QCD ($N_c = 3$) correspond to the small
quark mass limit.  We  find pion nucleon coupling
strengths are reduced to order ${\cal O}(1)$ rather than ${\cal O}(N_c)$.  Under the
assumption that in the large $N_c$ limit the sigma meson mass is larger
than that of the omega, and that the omega-nucleon coupling constant is larger
than that of the sigma, we argue that the nucleon-nucleon large range potential
is weakly attractive and admits an interaction energy of order
$\Lambda_{\rm QCD}/N_c^{5/2} \sim 10$ MeV.  With these assumptions on coupling and
masses, there is no strong long range attractive channel for nucleon-nucleon
interactions, so that nuclear matter at densities much smaller than that where
nucleons strongly interact is a weakly interacting configuration of nucleons
with strongly  interacting localized cores.  This situation is unlike the case
in the conventional large $N_c$ limit, where nuclear matter is bound with
binding energies of order the nucleon mass and forms a Skyrme crystal.

\end{abstract}

\maketitle

\section{Introduction}

It is standard lore that the large $N_c$ limit predicts a vacuum expectation
value for the sigma field, $f_\pi$ which scales as $\sqrt{N_c}$, and that
baryons have a mass of order $N_c$~\cite{Witten:1979kh}.  A model for baryons
is provided by the Skyrme model; within the Skyrme model, the meson nucleon
interactions are strong, of order $N_c$~\cite{Adkins:1983ya}.  This includes
pion-nucleon interactions, and for massless pions leads to very strong long
range forces~\cite{Witten:1979kh}.

The long range force associated with pion exchange leads to the conclusion that
the binding energy of nuclear matter
is of order the nucleon mass, and that a nucleon liquid would be unstable,
forming a Skyrme crystal of tightly bound nucleons~\cite{Klebanov:1985qi}.  In
nature, nuclear matter is weakly bound, with binding energy of order $15$~MeV.
In spite of the great successes of large $N_c$ phenomenology for mesons, it
appears that the large $N_c$ limit dramatically fails to describe the most
basic feature of nuclear matter, that it is weakly bound~\cite{Hidaka:2010ph}.

The large strength of meson forces is impossible to evade in the standard
Skyrme model of nucleons.  The Skyrme model describes a two-baryon solution,
and as a function of position, the energy scales as $N_c$.  This has led some
to modify the Skyrme model, by ignoring the kinetic energy term in the
non-linear sigma model~\cite{Adam:2010ds}.  It is then argued that
the Wess-Zumino-Witten term~\cite{Wess:1971yu,Witten:1983tw} generates a
baryon-baryon interaction, and this combined with the mass term for
the pions generates stable skyrmions.  Remarkably, such skyrmions saturate the
BPS bound that means to leading order in large $N_c$, they are non-interacting.
The essential feature of such consideration is that the ordinary
sigma model kinetic energy term vanishes  corresponding to vanishing vacuum
expectation values for  the sigma field.

While the BPS Skyrme model has the attractive feature that the interaction
energy of nucleons is generically
weak~\cite{Bonenfant:2010ab,Adam:2011zz},
it has the unattractive feature that the baryon number density squared is
stabilized by the mass term associated with explicit breaking of chiral
symmetry.   As we shall see, this results in the baryon density being of order 
$N_c m_\pi$, and the massless pion limit of the theory corresponds to a baryon
of infinite extent. We can see this from elementary scaling arguments.  The
baryon self  interaction associated with omega meson exchange is of order
\begin{equation}
  \int d^3x ~{N_c \over M_\omega^2 }~\rho_B(x)^2 \sim {N_c \over {M_B^2R^3}}, 
\end{equation}
where $R$ is the nucleon radius.  The stabilizing mass generating by the ``sigma term'' is of order 
\begin{equation}
  \sim N_c m_q \Lambda_{\rm QCD}^3 R^3 .
\end{equation}
Extremization gives
\begin{equation}
   R_{\rm nuc} \sim m_q^{-1/6} \Lambda_{\rm QCD}^{-5/6} 
\end{equation}
and the nucleon mass of order
\begin{equation}
   M_{\rm nuc} \sim N_c m_q^{1/2} \Lambda_{\rm QCD}^{1/2} \sim N_c m_\pi\,.
\end{equation}
The baryon number density is of order 
\begin{equation}
1/R^3 \sim m_q^{1/2} \Lambda^{5/2} \sim m_{\pi} \Lambda_{\rm QCD}^2\,.
\end{equation}

The vanishing of the nucleon mass in the chiral limit is an unfortunate
consequence of this theory.
Nevertheless, the idea that the vacuum expectation value of the sigma field is small, and as well the corresponding kinetic term in the non-linear sigma model, might have merit and we will consider 
this assumption in this paper.  We will see that if we assume the kinetic term is suppressed by one order of $N_c$,
we find a different dependence on quark mass for the nucleon mass and radius, and that for the values
for which there is lattice data from DeGrand and Liu~\cite{DeGrand:2016pur},
the resulting nucleon mass is consistent with the data.

A vanishing vacuum expectation value for the scalar field in leading order in
$N_c$ corresponds to a vanishing mass for the sigma particle in this 
order.  A vanishing mass for the scalar sigma field does happen in the large 
$N_c$ limit of QCD in two dimensions~\cite{tHooft:1974pnl}.  It is also not so implausible in four
dimensions.  In leading order in $N_c$, the four dimensional theory is a
non-interacting theory.  With a negative mass squared for the sigma field, this
large $N_c$ theory would be unstable, since to leading order in $N_c$,
interaction terms which would stabilize the theory vanish.  A positive mass
squared term would not allow for chiral symmetry breaking.  Assuming a negative
mass squared term generated in
non-leading order in $N_c$ would allow for a stable theory in the strict large
$N_c$ limit.  The potential would be flat in this limit, allowing for symmetry
breaking as required by the Coleman-Witten theorem~\cite{Coleman:1980mx}.  The
value of the condensate would be determined by the next to leading order
corrections to the sigma model.

It is difficult to provide an explicit mechanism for how the symmetry breaking
negative mass squared term might vanish.  It is interesting that the low energy
sector that results for massless quarks is scale invariant at the mean field
level for the effective pion-sigma action.  Scale invariance is associated with
critical phenomena. Perhaps such scale invariance might be argued by
renormalization group methods in the large $N_c$ limit.  As we will see, this
scale invariance argument is also useful for arguing the form of the effective
action that generate the nucleon mass.
In any case, we will take throughout this paper an unproven assumption that
this masslessness exists and shall explore the consequences of this assumption.  

One can ask whether this $1/N_c$ behavior of the squared mass of the sigma
field is preserved by higher order radiative corrections.  This is equivalent
to asking if  the $1/N_c$ behavior is natural.  Indeed, when one computes
the radiative correction to the sigma mass, the lowest order correction arises
from a tadpole diagram.
The four meson interaction in the diagram is of order $1/N_c$ and the overall
quadratic divergence is cutoff
at the QCD scale so this gives a contribution to the mass squared of order
$\Lambda_{\rm QCD}^2/N_c$, and indeed the behavior of the scalar mass term is
natural.  Higher order corrections indeed maintain this behavior.  

Another implication of these results is that $f_\pi \sim {\cal O}(1)$ in powers of
$N_c$.  On the other hand standard counting in powers of $N_c$ using quark
counting gives $f_\pi \sim \sqrt{N_c}$. In the considerations below, we will limit
ourselves to momentum scales which are less than $\Lambda_{\rm
QCD}/\sqrt{N_c}$.  If for example we compute the matrix elements of the vector
current squared, it is given by a vacuum polarization diagram involving a quark
loop,
and this appears to be of order $N_c$ corresponding to  a current with a
typical value of order $\sqrt{N_c}$.  However, if we look at low momentum
scales, of order $q \sim \Lambda_{\rm QCD}/\sqrt{N_c}$, the contribution to the
quark loop involving gluons must be summed.  The leading order contribution in
powers of $N_c$ are one vector meson states such as the $\rho$ meson, and
this state decouples at zero momentum transfer as $q^2/M_{\rho}^2$, suppressing
the contribution by order $1/N_c$.  Now if a  vector meson coupling to this
channel shrinks to zero mass in the large $N_c$ limit, then the ordinary
counting can be maintained.  

We shall later argue that the sigma meson  mass becomes small  
in the large $N_c$ limit.  If this is the case, the ordinary $N_c$
counting should work for vector current matrix elements.
The axial vector channel is however different.  In the axial isoscalar vector
channel there is a U(1) anomaly, and we will argue this will result in a pseudo-scalar particle with mass of order $\Lambda_{\rm QCD}$.  

Let us consider the two axial vector current correlation function.  This correlation function involves a quark-antiquark pair, the interactions of which can lead to a pole corresponding to an axial vector meson We   
assume that, as is the case for the scalar axial vector mesons also have masses of order $\Lambda_{\rm QCD}$. 
In this case,  axial vector
meson contributions will be suppressed at small $q^2$.  This leaves two meson
intermediate and the pion  states which are suppressed by powers of $N_c$.
Note that the contribution from the scalar sigma and pion to the axial vector
isotriplet current can be written in terms of the low momentum degrees of
freedom as
\begin{equation}
  J_{\mu a 5} \sim \sigma \{ {\stackrel{\rightarrow} \partial_\mu- \stackrel{\leftarrow} \partial_\mu} \}\pi^a\,.
\end{equation}
For the cases where $f_\pi = \langle\sigma\rangle \sim {\cal O}(1)$ in powers of $N_c$, this
current has matrix elements of order 1,
consistent with the reasoning above.  A necessary condition for this to be
maintained, therefore, is that the isovector axial vector mesons have a mass of
order $\Lambda_{\rm QCD}$. 

We can easily see why the $\eta^\prime$ mass is of order one in our scenario.  It follows simply by the assumption that the energy density dependence of the $\theta$ angle of QCD is of order one because 
\begin{equation}
       m^2_{\eta^\prime} \sim {1 \over f_\pi^2} {{d^2E} \over {d\theta^2}} \sim \Lambda^2_{\rm QCD}\,,
\end{equation}
where we have assumed the $U(1)$ current is of the form
\begin{equation}
J_{\mu 5} \sim \sigma \{ {\stackrel{\rightarrow} \partial_\mu- \stackrel{\leftarrow} \partial_\mu} \}\eta^\prime\,.
\end{equation}
Since the $U(1)$ axial symmetry is explicitly broken, there is no spontaneous
symmetry breaking so the parity doublet of the pion, the scalar isovector
mesons presumable have masses of order $\Lambda_{\rm QCD}$.

There is no reasonable theory of nuclear matter if the scale $\sigma$ becomes
massless and there is no other small mass iso-singlet vector, since this
generates an attractive force on baryons of order $N_c$ and will cause nuclear
matter to collapse.  One needs also that the $\omega$ meson becomes massless
also, with $m_\omega \le m_\sigma$ and that in this limit, $\omega$ nucleon
coupling is larger than that of the
$\sigma$ meson.  This guarantees that in the isosinglet channel at distances
less than $R \le \sqrt{N_c}/ \Lambda_{\rm QCD}$, the over all isosinglet force
is repulsive.  In various theories such as those with a hidden local gauge
symmetry, the omega meson mass
is of order $f_\pi/\sqrt{N_c}$.  We will need to have this possibility here as well.
Presumably there is a similar behavior for the $\rho $ meson.  So the theory we
propose would have a Goldstone pion, a small mass sigma and omega,
$m_\omega \le  m_\sigma \sim 1/\sqrt{N_c}$.  We will make such assumptions in
this paper, and explore the consequences.

In the next section, we write out the explicit action for the pion-nucleon
sigma model.  We then use this action to compute $f_{\pi}$, the sigma meson
mass and the pion mass.  We compute these quantities as a function of the quark
mass, the number of colors $N_c$, the four meson coupling strength $\lambda$ and the negative mass
squared which  drive the symmetry breaking.  We show that our results are in
good accord with those of DeGrand and Liu for $N_c =3$ and determine the values
of the underlying parameters of our sigma model~\cite{DeGrand:2016pur}.  The
values for quark masses in the computation of Bali et. al. are quite large,
and they work in the quenched approximation, and we do not compare
with their data~\cite{Bali:2013kia},
We find there are two possible cases:  In the first the quark mass is
generically $m_q \gg  \Lambda_{\rm QCD}/N_c^{3/2}$.
In this limit, symmetry breaking is driven by the explicit symmetry breaking of the quark mass.  The large $N_c$ counting is conventional for physical quantities.  The data of DeGrand and Liu is in this range~\cite{DeGrand:2016pur}.  The other case is
$m_q \ll \Lambda_{\rm QCD}/N_c^{3/2}$. 
In this range physical quantities have unanticipated dependences on $N_c$
and there is no Monte Carlo data for $N_c$ dependence for this case.  For  QCD,
$(N_c = 3)$,  the physical value of quark masses is in this region.

In the fourth section we compare the sigma model action prediction to the
computations of DeGrand and Liu for variable $N_c$~\cite{DeGrand:2016pur}.  We
find fair agreement with their results, although the quark masses considered
are sufficiently large   so that deviations from the naive large $N_c$ scaling
predictions are small.

In the fourth section, we consider baryons.  We argue that the baryon mass for
the small quark mass limit scales as
$M_B \sim \sqrt{N_c} \Lambda_{\rm QCD}$, and that the radius $R \sim \sqrt{N_c}/\Lambda_{\rm QCD}$.  These scaling relation imply that the baryon a mass of order $\sqrt{N_c} f_\pi$, and this
dependence is consistent with  the data of DeGrand and Liu for large quark
masses~\cite{DeGrand:2016pur}.  The  data available are not in the small quark
mass limit where the unexpected behaviour as a function of $N_c$ is found.  Nevertheless, the agreement with our $N_c$ dependence is only fair for the range of 
parameters considered by DeGrand and Liu.

In the last section, we consider general properties of the nucleon-nucleon
force.  We find the pion to be couple not with a coupling of order $\sqrt{N_c}$
to the nucleon but of order 1.  The pion force generates a long range tail for
nucleon-nucleon interaction. 

To achieve a theory of the nucleon that has a reasonable small interaction strength for nuclear matter, we needed to make  drastic assumptions about the behaviour of QCD in the large $N_c$ limit.  In the future, we intend to investigate various
Lagrangian for scalar and vector mesons to see under what set of assumptions, if any, such a description is 
internally consistent.  It is also true that the behaviour we predict makes testable predictions for 
Monte-Carlo simulations of the large $N_c$ limit of QCD.  Unfortunately, the restriction to very small mass quarks makes explicit computation very difficult.  

\section{The Low Energy Linear Pion Sigma Model as a Function of $N_c$}
\label{Sect:Model}

We begin with the effective action
\begin{equation}
	\label{Eq:S}
	S = {1 \over 2}\left[(\partial \phi)^2 + (\partial \pi)^2\right]  - {1 \over 2} {m^2 \over N_c} (\phi^2 + \pi^2) + {\lambda \over {4N_c}} (\phi^2 + \pi^2)^2 - \sqrt{N_c} m_q \mu^2 \phi. 
\end{equation}
 Here $\phi$ is the sigma field and $\pi$ is the pion field.  The quark mass is
 $m_q= (m_u+m_d)/2 \sim 3.5$~MeV.
 The parameters $m, \lambda, \mu$ will be determined from the sigma meson mass,
 the pion decay constant $f_\pi$ which is the vacuum expectation value of
 $\phi$ and by the pion mass.
 
 We are here assuming that the negative mass squared term in our action is of
 order $1/N_c$.   The conventional large $N_c$ limit is obtained if
 $m^2 \sim N_c$.  The $\sqrt{N_c}$ in front of the last term (the sigma term)
 is of the correct order for the large $N_c$ limit.
 
 The vacuum expectation values of the sigma field is determined by the roots of
 \begin{equation}
	 \label{Eq:phi} 
 \phi^3 - \phi ~{m^2/\lambda}  - N_c^{3/2} m_q ~\mu^2/\lambda = 0\,.
 \end{equation}
 We denote this root by $\phi_0$.  
 
 One can solve for the values of the parameters $m^2$, $\mu^2$ and $\lambda$ in
 terms of the masses of the sigma meson, the pion and the expectation value of
 the scalar field $\phi_0$.  The vacuum values for $N_c =3$ are taken to be~
 \begin{equation}
	 m_\pi = 140~\text{MeV}, \quad  \phi_0 \sim f_\pi \sim 130~\text{MeV}, \quad 
 	 M_\sigma \sim 950-1050~\text{MeV} . 
	 \label{Eq:Input}
 \end{equation}
 The solutions are
 \begin{equation}
   \lambda = N_c {{m_\sigma^2 - m_\pi^2} \over {2\phi_0^2}},
 \end{equation}
 \begin{equation}
	 \label{Eq:mu2}
   \mu^2 = {{m_\pi^2 \phi_0} \over {m_q \sqrt{N_c}}},
 \end{equation}
 \begin{equation}
   m^2 = N_c {{m_\sigma^2 - 3 m_\pi^2} \over 2}. 
 \end{equation}
Note that if we were to take $m^2 \sim N_c$ for the conventional large $N_c$
limit, we would have $\phi_o \sim \sqrt{N_c}$, and the values of $\lambda$,  $
\mu^2$ and $m^2$ would be $N_c$ independent.  Our philosophy will be to take
these values from the case $N_c = 3$, fix them, and then determine the 
variation of $m_\pi$, $m_\sigma$ and $\phi_0 = f_\pi$ predicted by the variation of $N_c$.

Notice there are two regimes for solution to this equation.  We will get these
values more precisely later, but for now
if we assume that $\mu \sim m \sim \Lambda_{\rm QCD}$ and $\lambda \sim 1$,
these regimes are
the small and large quark mass limits which are separated by $m_q \sim
\Lambda/N_c^{3/2}$.
When we put in properly determined numbers, we will see that the small quark
mass regime is characteristic of the physical values for the
real world of pions, sigma mesons and $f_\pi$.  However, we will see that for
almost all of the range covered in the work of DeGrand, the large quark mass
region dominates.  Notice that this also implies  that as $N_c \rightarrow
\infty$, there is a non-uniformity of the small quark mass limit.  The small
mass limit is what is required for good phenomenology, so we must always take
$N_c$ large but finite and consider masses in the range $m_q \ll
\Lambda/N_c^{3/2}$ for the correct physical limit.

Let us explore solutions in these different limits. 
In the small mass limit (Region I), the VEV is determined by ignoring the sigma
term (term proportional to the quark mass)
\begin{equation}
	\label{Eq:VEV1}
  \phi_0 = m/\sqrt{\lambda} .
\end{equation}
Because of the extra $1/N_c$ in the mass term, the VEV for case one is $N_c$ independent.

In the large quark mass limit (Region II)
\begin{equation}
	\label{Eq:VEV2}
\phi_0 = \sqrt{N_c} (m_q\mu^2/\lambda)^{1/3} .
\end{equation}
In this latter case, the scalar field has the canonical dependence on $N_c$ as
expected in the ordinary  large $N_c$ limit.  We will see that this will
guarantee the expected large $N_c$ behavior for physical quantities in this
limit.

We can now compute the sigma mass:
\begin{equation}
  m_\sigma^2 = - {m^2 \over N_c} + {{3\lambda} \over N_c} \phi_0^2. 
\end{equation}
For region 1,
\begin{equation}
m_\sigma^2 = {{2m^2} \over {N_c}}. 
\end{equation}
In this small mass region, the sigma mass shrinks to zero as $N_c \rightarrow
\infty$.  Of course for any fixed quark mass, we are only in the region up till
some large $N$ and then we move into the effective large mass region. In Region
II
\begin{equation}
  m_\sigma^2 = 3 (m_q\mu^2 \sqrt{\lambda})^{2/3} .
\end{equation} 

The pion  mass can also be computed in these two limits:

Region I
\begin{equation}
m_\pi^2 = \sqrt{N_c} \lambda m_q {\mu^2 \over m^2}
\end{equation}
and 
Region II
\begin{equation}
m_\pi^2 = (m_q\mu^2\sqrt{\lambda})^{2/3}. 
\end{equation}

Let us now determine parameters for the world we live in\footnote{For certainty
we use $M_\sigma = 1000$~MeV.}  
, see Eqs.~\eqref{Eq:Input}:
\begin{equation}
	m  =  1112.3~\text{MeV}, \quad \mu =  648.3~\text{MeV}, \quad \lambda = 76.7. 
\end{equation}
The obtained value of the coupling constant is somewhat large even if we take into account that 
the expansion parameter that controls perturbative computations is $\lambda/(4 \pi^2 N_c) \approx 0.7$.
This either means that the model at hand is too simplistic or that the lattice artifacts 
dominate in the observables. Here our strategy is to be pragmatic and to take the fitted 
values at their face value. 

The edge of the region where the large limit dominates is when the quartic term
in the potential dominates over the quadratic.  This is when
\begin{equation}
  \lambda \phi^2 \sim m^2
\end{equation}
or when
\begin{equation}
	\label{Eq:mqCr}
	m_q \sim m {m^2 \over \mu^2} {1 \over {\sqrt{\lambda}} }{1 \over {N_c^{3/2}}} = 72~\text{MeV}.
\end{equation}


\section{Mesons at $N_c = 3$: Comparison to LQCD Results}

\begin{figure}[ht]
	\centerline{
		\includegraphics[width=0.48\linewidth]{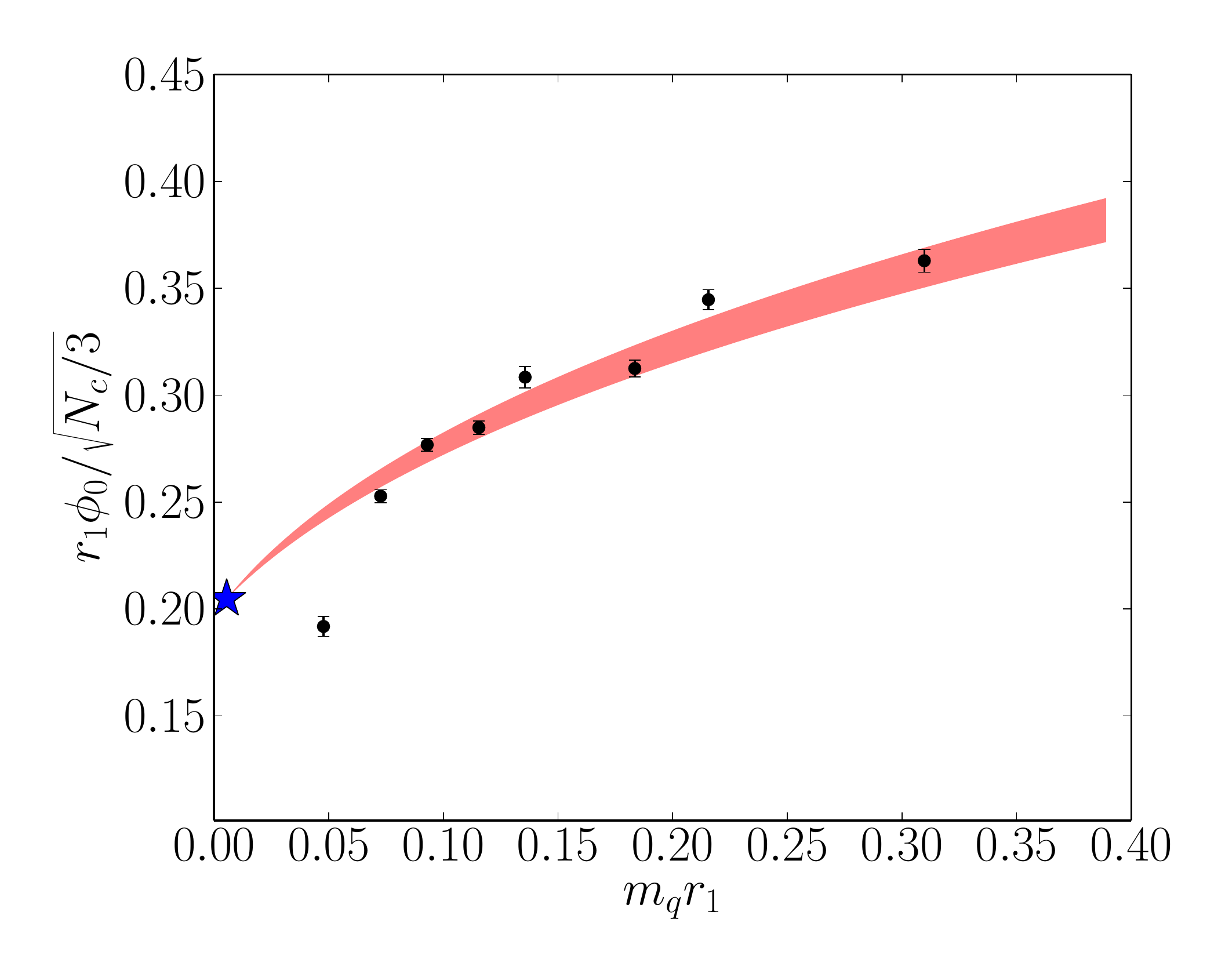}
		\includegraphics[width=0.48\linewidth]{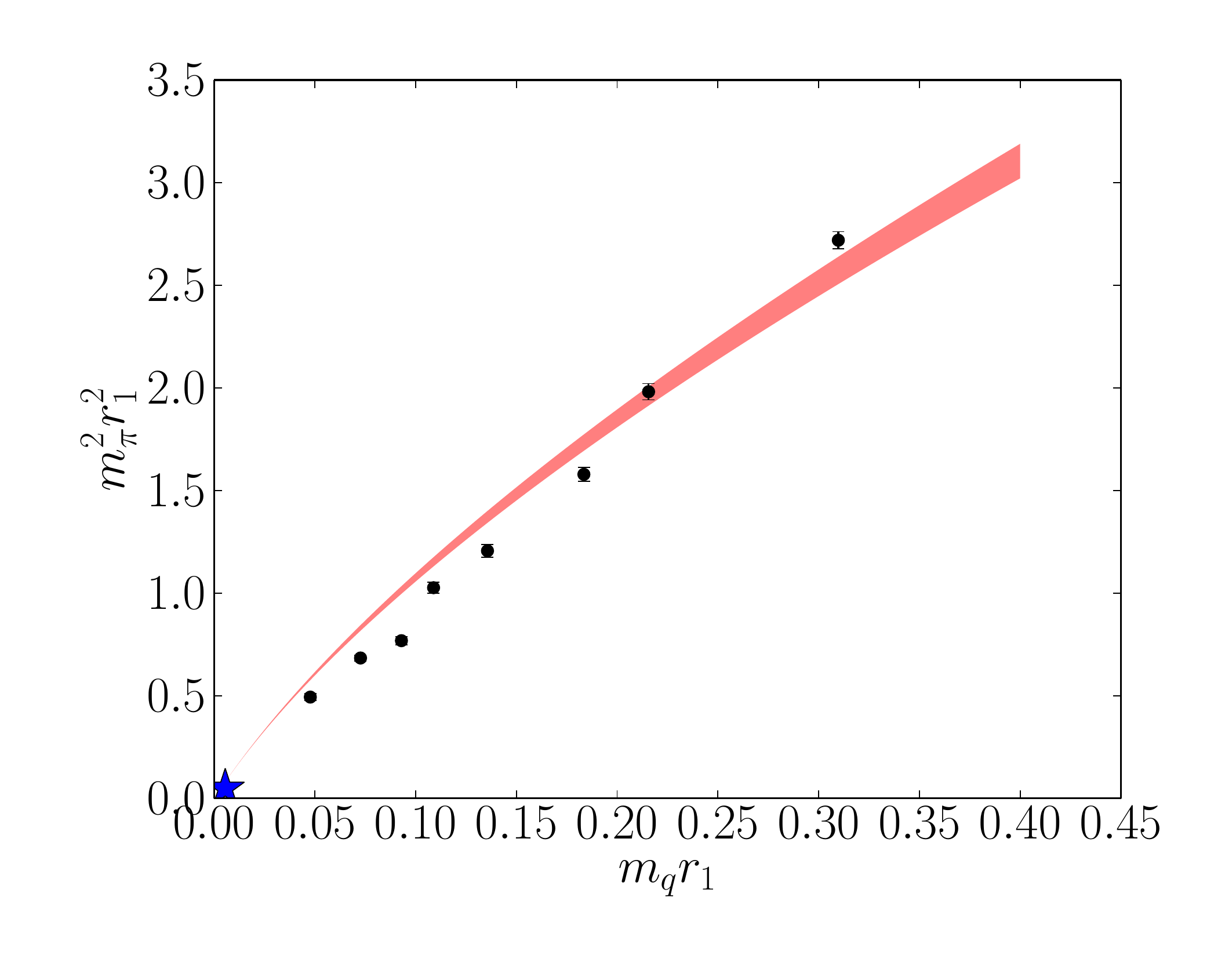}
}
\caption{ The pion decay constant (left panel)  and the pion mass (right panel) as a function of the quark mass for 
	$N_c=3$. The shaded region corresponds to the variation of the sigma mass in the range from 950 to 1050 MeV. 
	The symbols show  the lattice data  from Ref.~\cite{DeGrand:2016pur}. 
	The stars show the physical values of the observables.  
}
\label{fig:PS}
\end{figure}

Using the model described in Sect.~\ref{Sect:Model}
we attempt to fit the dependence of the mesonic properties on the quark mass 
obtained from the lattice data~\cite{DeGrand:2016pur} for $N_c=3$. 
We would like to point out that the model  has a few parameters and fixing 
the parameters at the physical quark mass leaves us with only one parameter 
to vary -- the sigma mass. This over-constrains  the model and we fail to achieve 
a good fit of the lattice data. We thus allow small variation of the physical 
parameter $m_\pi$. This variation can be attributed to either 
systematics of the lattice calculations or, which is more plausible, to 
relevant physics our model fails to capture. 
We get a good description of the 
lattice if we increase $m_\pi$ to 180 MeV. 

In Figure~\ref{fig:PS}, we show the dependence of the pion decay constant and the 
pion mass
on the quark mass in the model; the results are compared to 
the lattice QCD calculations from Ref.~\cite{DeGrand:2016pur}. 
The model results are obtained by solving the stationarity conditions, Eq.~\eqref{Eq:phi}; 
the pion mass is then computed. 
The 
stars in Fig.~\ref{fig:PS}  denote the physical values of the observables. 

Although the model misses the first LQCD data point for the pion decay constant,  
the overall description of the lattice results 
is quite remarkable given the simplicity of the model. 

\begin{figure}[ht]
	\centerline{
		\includegraphics[width=0.48\linewidth]{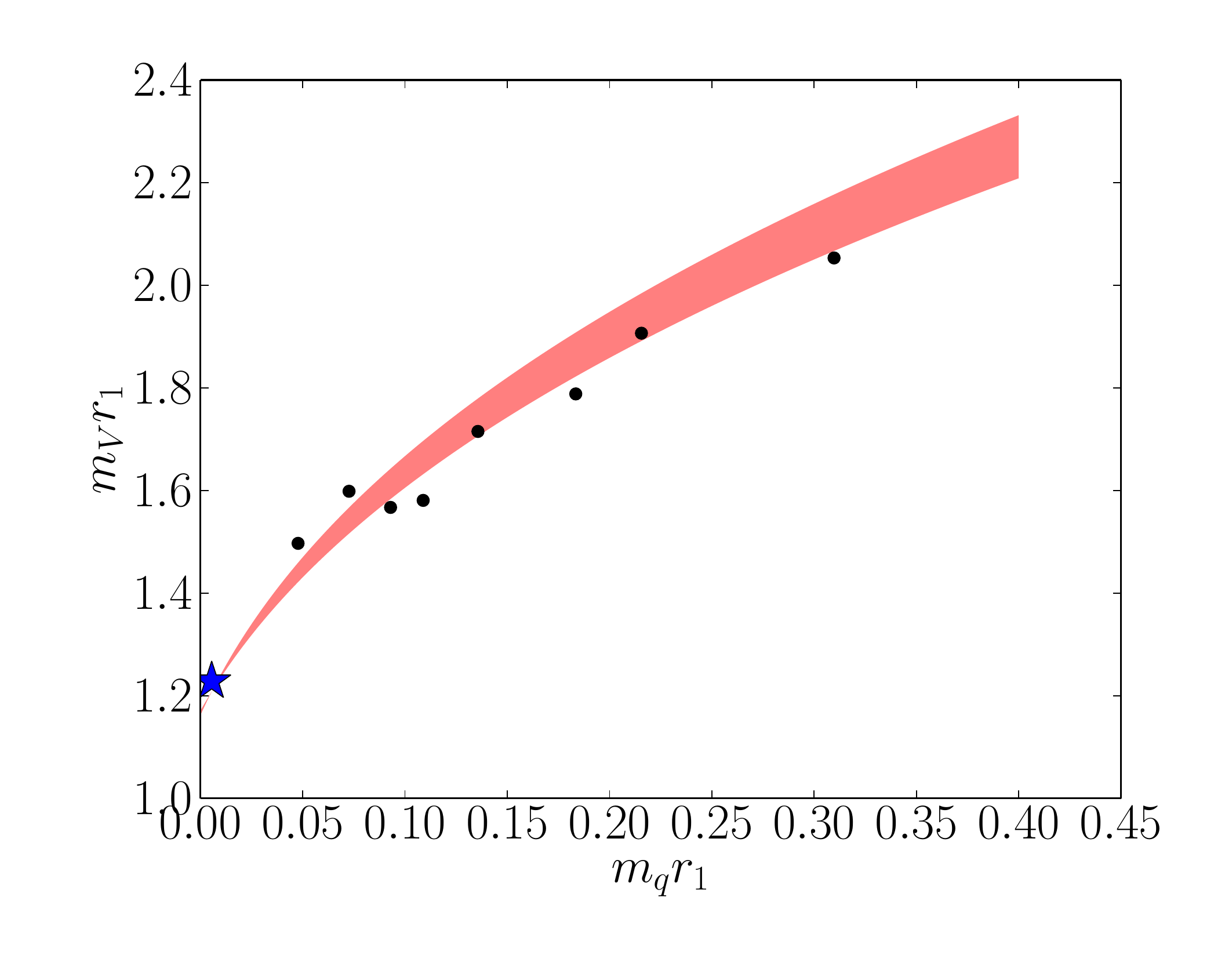}
}
\caption{ The vector meson mass as a function of the quark mass for 
	$N_c=3$. 
	The symbols show  the lattice data  from Ref.~\cite{DeGrand:2016pur}. 
	The stars show the physical values of the observables.  
}
\label{fig:rho}
\end{figure}

Additionally, we compare our results for the vector meson mass, assuming $m_V\propto f_\pi$, see
Fig.~\ref{fig:rho}. 

Using this fit, we can proceed by computing the observables for any number of colors.

\section{Variable $N_c$}
\begin{figure}
	\centerline{
		\includegraphics[width=0.48\linewidth]{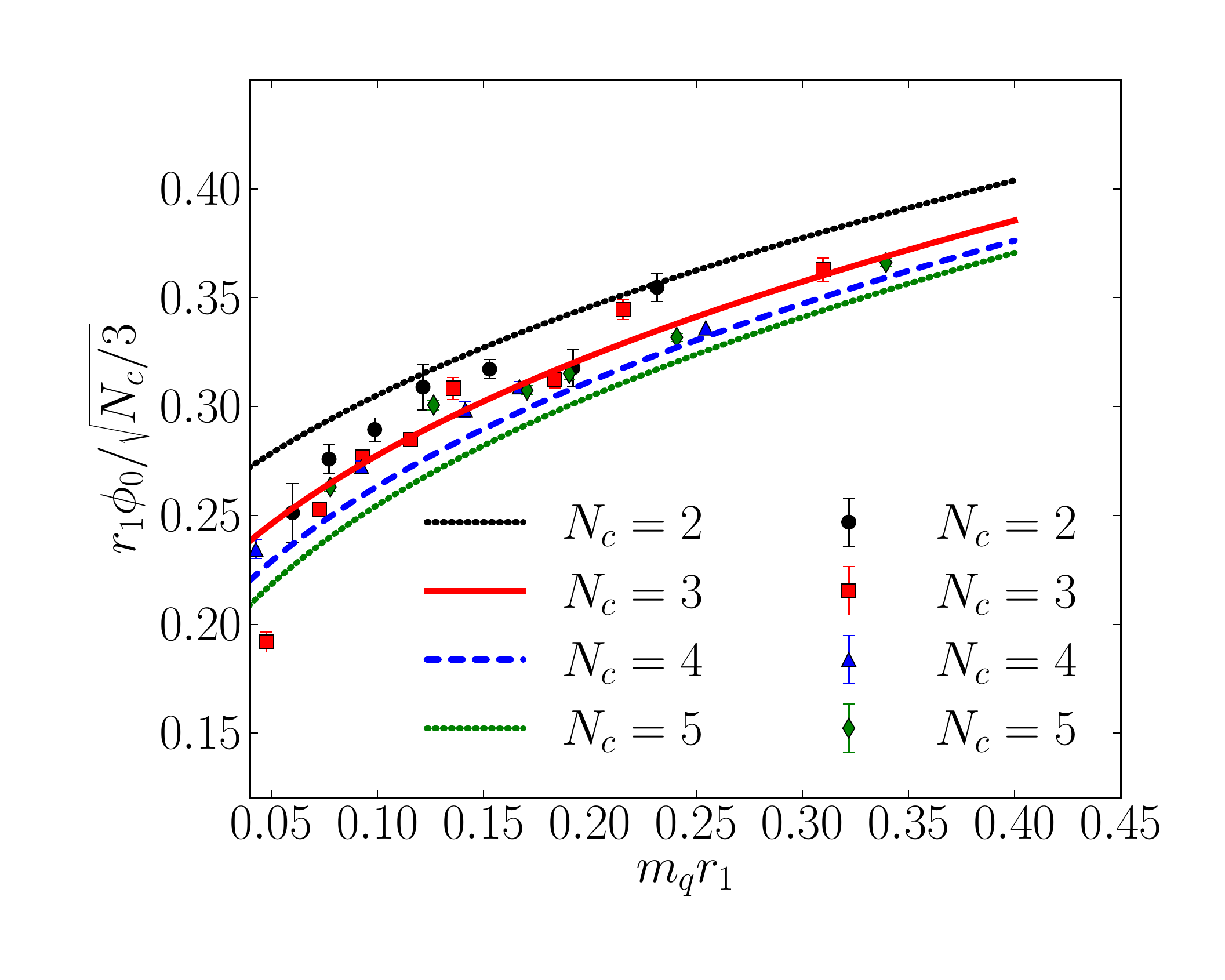}
}
\caption{ The pion decay constant as a function of the quark mass for 
	$N_c=2, 3, 4, 5, \infty$.
	The model results is shown by the curves. 
	The sigma mass in the model is fixed to 1000 MeV.
	The lattice data (symbols) is from Ref.~\cite{DeGrand:2016pur}. 
}
\label{fig:PS_Nc}
\end{figure}

Recent calculations by DeGrand and Liu show the dependence of various observables on $N_c$~\cite{DeGrand:2016pur}. 
Here we want to demonstrate that, in their range of the quark masses (in the model, it is Region II described 
in Sect.~\ref{Sect:Model}), our model 
follows the data and exhibits  the conventional dependence on the number of
colors. 
In Fig.~\ref{fig:PS_Nc} we show the pion decay coupling constant as a function of the mass  for various $N_c$. 
For the model, we also show the limit of $N_c\to \infty$. As 
seen from the figure, for these values of the quark masses the large $N_c$ 
limit describes lower values $N_c$ quite well, which demonstrates the conventional scaling of Region II, see Eq.~\eqref{Eq:VEV2}. 

The important question this exercise allows us to address is: Are the lattice
data in a range where we can 
rule out our $1/N_c$ hypothesis for the negative mass-square term of the action~\eqref{Eq:S}? 
The answer is that no, we cannot rule this hypothesis out based on the {\it available} data.

\begin{figure}
	\centerline{
		\includegraphics[width=0.48\linewidth]{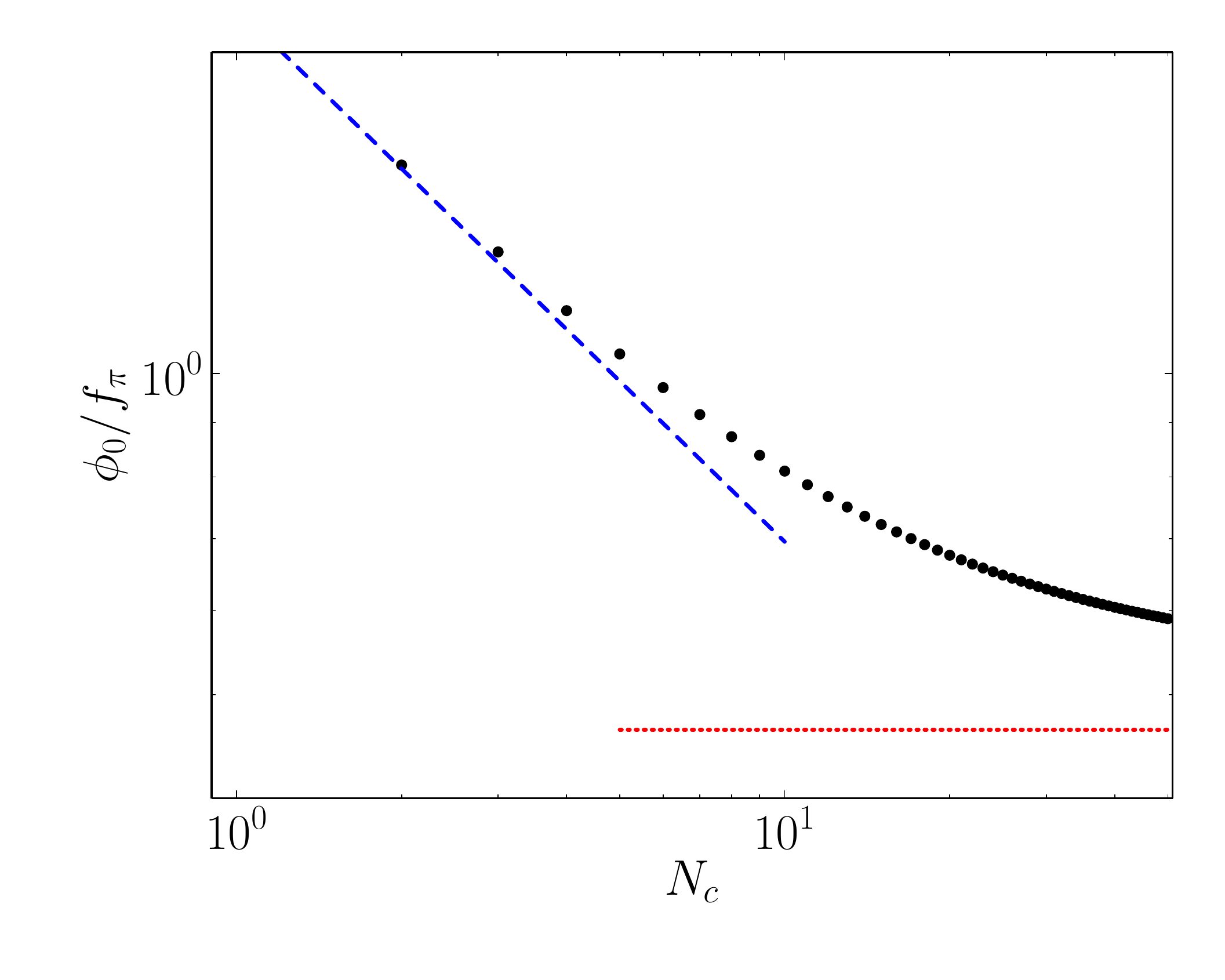}
}
\caption{Illustration of the different regions in approaching the large $N_c$ limit: the pion decay constant as 
	a function of $N_c$. The dashed line shows 
	the Region I, see Eq.~\eqref{Eq:VEV1}; the dotted line shows the Region II, see Eq.~\eqref{Eq:VEV2}.  
}
\label{fig:PS_regimes}
\end{figure}

As we argued in the introduction, the relevant region for the physical quark 
masses is Region 1. Indeed by applying the argument presented in Eq.~\eqref{Eq:mqCr} to the physical quark mass   
and computing the critical $N_c^{\rm cr}$ where the transition from Region I to Region II 
takes place, we get $N_c^{\rm cr} \approx 22$. We also demonstrate this in Fig.~\ref{fig:PS_regimes}, 
where the dependence of the pion decay coupling constant on the number of colors for the physical 
quark mass is plotted. The two asymptotic regimes, see  Eq.~\eqref{Eq:VEV1}
and  Eq.~\eqref{Eq:VEV2}, are also shown. This comparison shows that at the physical pion mass the relevant 
approach to the limit $N_c\to \infty$ should be considered in the case  of a small quark mass, Region I.

\section{Baryons}

Baryons in the large $N_c$ limit are conventionally assumed to be given by
non-perturbative solutions to the classical equation of motion of the
non-linear sigma model~\cite{Adkins:1983ya}.  These skyrmions have a topological
charge corresponding to baryon number.  Let us first consider general features
of such non-perturbative solutions.

\begin{figure}
	\centerline{
		\includegraphics[width=0.48\linewidth]{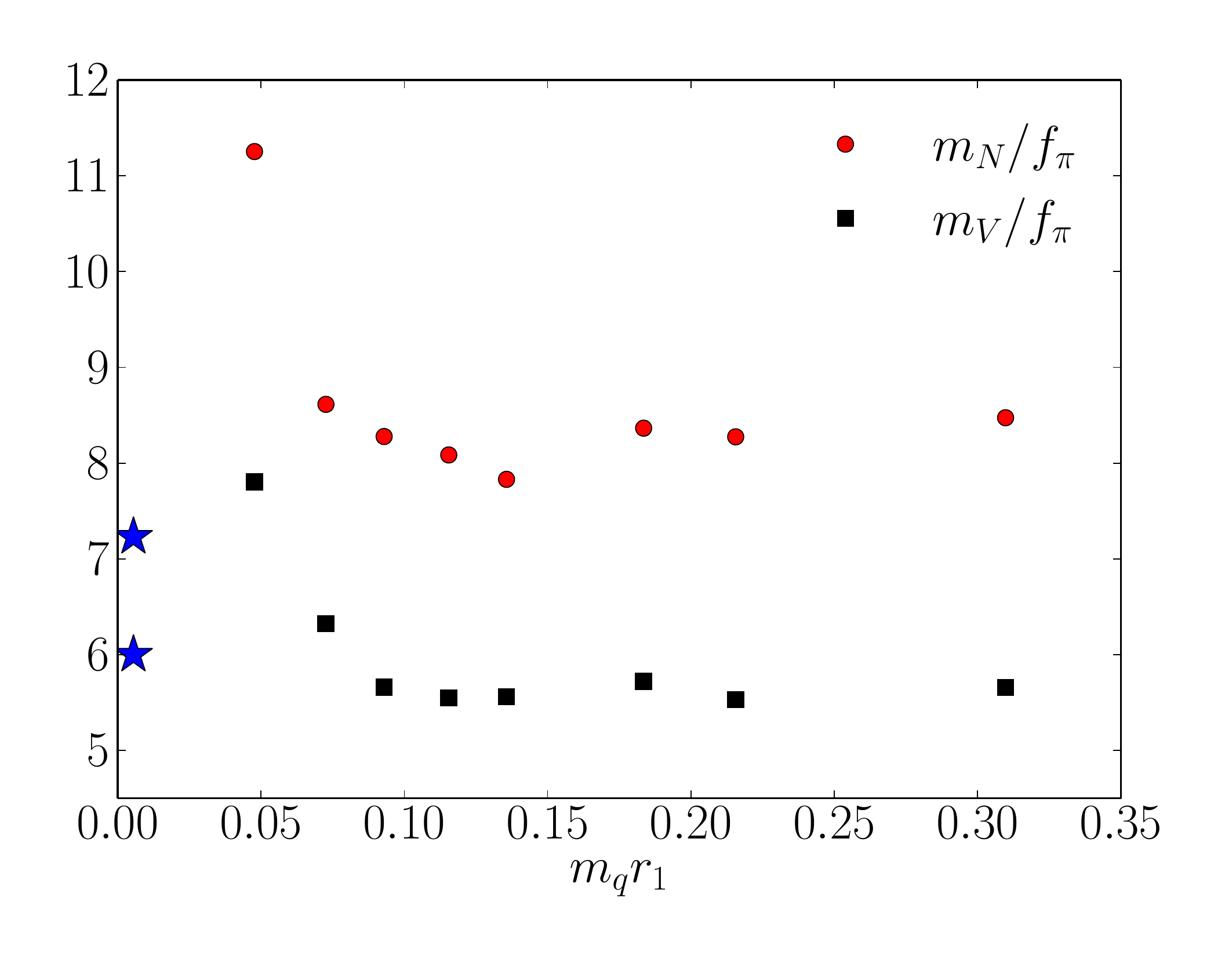}
}
\caption{The ratios of the nucleon and vector meson mass to the pion coupling constant
	as a function of the quark mass. This is representation of the results by  DeGrand and Liu, 
	see Ref.~\cite{DeGrand:2016pur}.
}
\label{fig:repr}
\end{figure}

In the small quark mass region where the large $N_c$ behaviour is
unconventional,  the potential is generically of order $1/N_c$.  Therefore the
kinetic energy term is of the order of potential energy if $R \sim \sqrt{N_c}/
\Lambda_{\rm QCD}$.  This implies the mass of the skyrmion is of order $M
\sim\sqrt{N_c} \Lambda_{\rm QCD}$.  More generally, even in the large $N_c$
region, the kinetic energy will trade off against the potential energy when $R
\sim \sqrt{N_c} /f_\pi$, and 
$M \sim \sqrt{N_c} f_\pi$.  
The lattice data seems to support it, see our representation of results by 
DeGrand and Liu in Fig.~\ref{fig:repr}. 
For the conventional Skyrme model treatment, this
would imply $M \sim N_c \Lambda_{\rm QCD}$ and $R \sim 1/\Lambda_{\rm QCD}$.

For large $m_{q}$, we would indeed find the correct $N_c$ counting for the mass
and radius of any solitonic solution, however the dependence upon quark mass is
nontrivial.  As the scale is set by $f_\pi$, for large quark mass we have the
radius
\begin{equation}
  R \sim \sqrt{N_c}/\phi_0 \sim (\lambda/m_q\mu^2)^{-1/3} .
\end{equation}
In order to be in the large quark mass limit appropriate for this expression,
we must have $R \ll \sqrt{N_c} /\Lambda_{\rm QCD}$.  Correspondingly, the mass
is
$M \gg \sqrt{N_c} \Lambda_{\rm QCD}$.  

In general, the only scale in our theory of pions and sigma mesons is $f_{\pi}$
so if there is a non-perturbative skyrmionic solution for the theory, its size
will generically be of order $\sqrt{N_c}/f_\pi$ and mass of order $\sqrt{N_c}
f_\pi$.  We might have thought that a derivative expansion for an effective
Lagrangian for the skyrmion would be well behaved because of the large
distances involved.  This is not the case as the mass scale associated
with this derivative expansion is the $\sigma$ mass, which is of this same size.

A light mass sigma meson will cause problems for baryonic matter.  The sigma
meson generates an attractive self interaction.  This will show up in a
skyrmion solution for two particle interactions that will have a long distance
attractive interaction that will generate a binding energy for the two skyrmion
solution that is of the order
of the skyrmion mass.  The sigma mass cuts this interaction off at large
distances, so large binding is generated at the size scale of the skyrmion and
is a hard core attraction.  In order to have a sensible theory, one needs a
compensating repulsion.  This might be given by the omega meson.  The coupling
of the omega to the nucleon would need to be larger than that of the sigma.
This is phenomenologically the case.  More importantly
the omega mass would have to shrink to zero like the omega in the large $N_c$ limit,
and $m_\omega \le m_\sigma$.
In models with a  hidden local gauge symmetry, this is plausible, since the
omega meson's mass is proportional to
$f_\pi/\sqrt{N_c}$.  It is also not so implausible if the large $N_c$ limit
corresponds to critical behaviour since both
the sigma meson and omega meson couple to isospin singlet density fluctuations.
With this added assumption
the omega and sigma mesons combine together to generate  a short distance
repulsive core.  If the typical separation of nucleon is large compared to this
core's size, it should not affect much the energy density of nuclear matter.
The hard core interaction will generate effects of order $\sqrt{N_c}
R^3_{\rm nucleon} \rho_{\rm baryon}$, so if the density of nuclear matter is
sufficiently low, the effects are small. This will be discussed more in the
next section when we discuss pion interactions.  At this point, we note that
the typical distance scale for the pion interactions is of order $1/m_{\pi}$,
and the since $R_{\rm nucleon}  m_\pi \sim \sqrt{N_c} m_{\pi}/\Lambda_{\rm QCD} \ll
N_c^{-1/4}$,
so that the effects of the hard core relative to the mass contributions to the
energy are suppressed  by at least a factor of $N_c^{-3/4}$.  In the case of
QCD for $N_c =3$, the bound that $m_q \le \Lambda_{\rm QCD}/N_c^{3/2}$
is satisfied with about an order of magnitude to spare, and for general $N_c$
there is always some sufficiently small quark mass where this will give an
acceptably small correction from the core.

\begin{figure}
	\centerline{
		\includegraphics[width=0.45\linewidth]{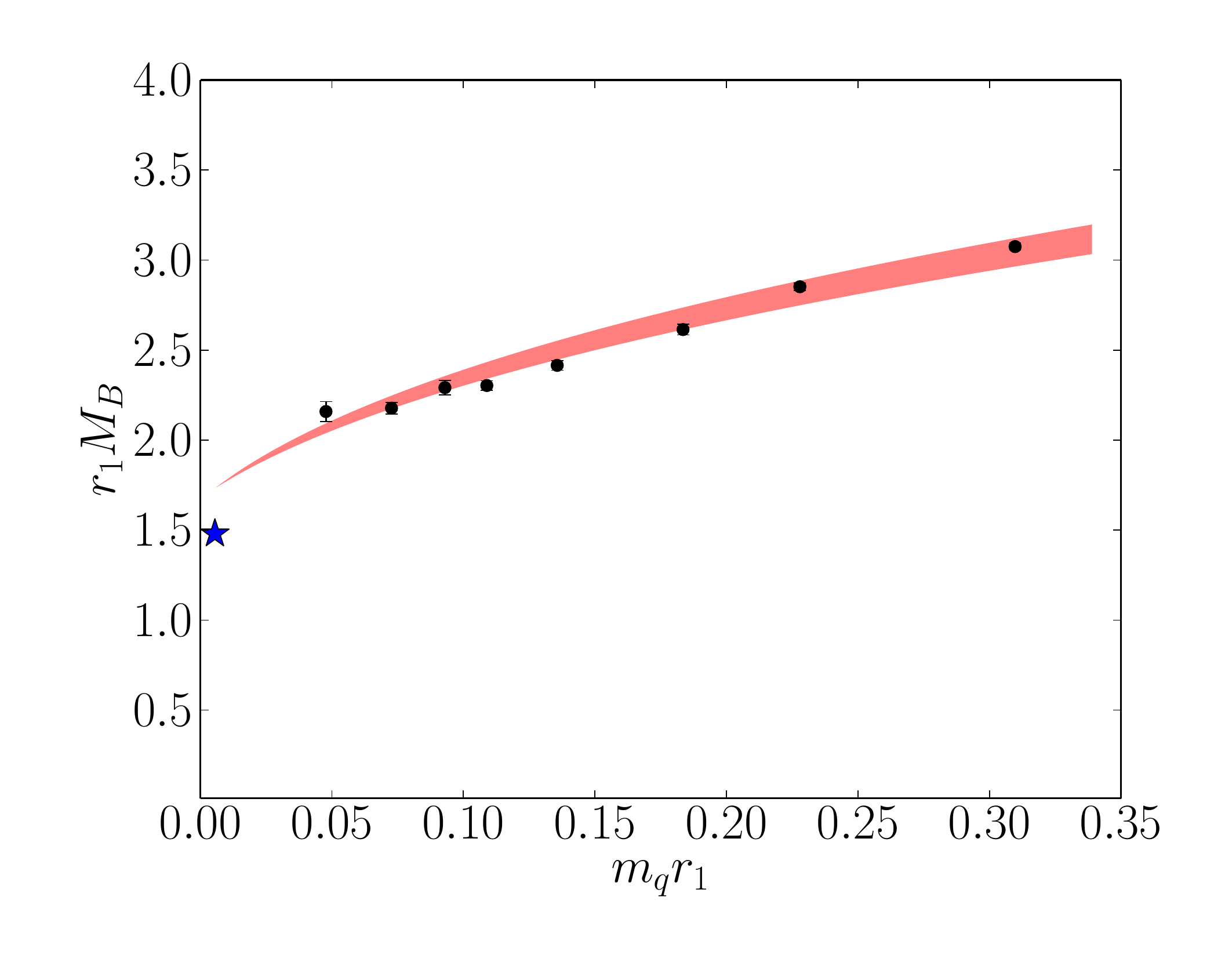}
		\includegraphics[width=0.45\linewidth]{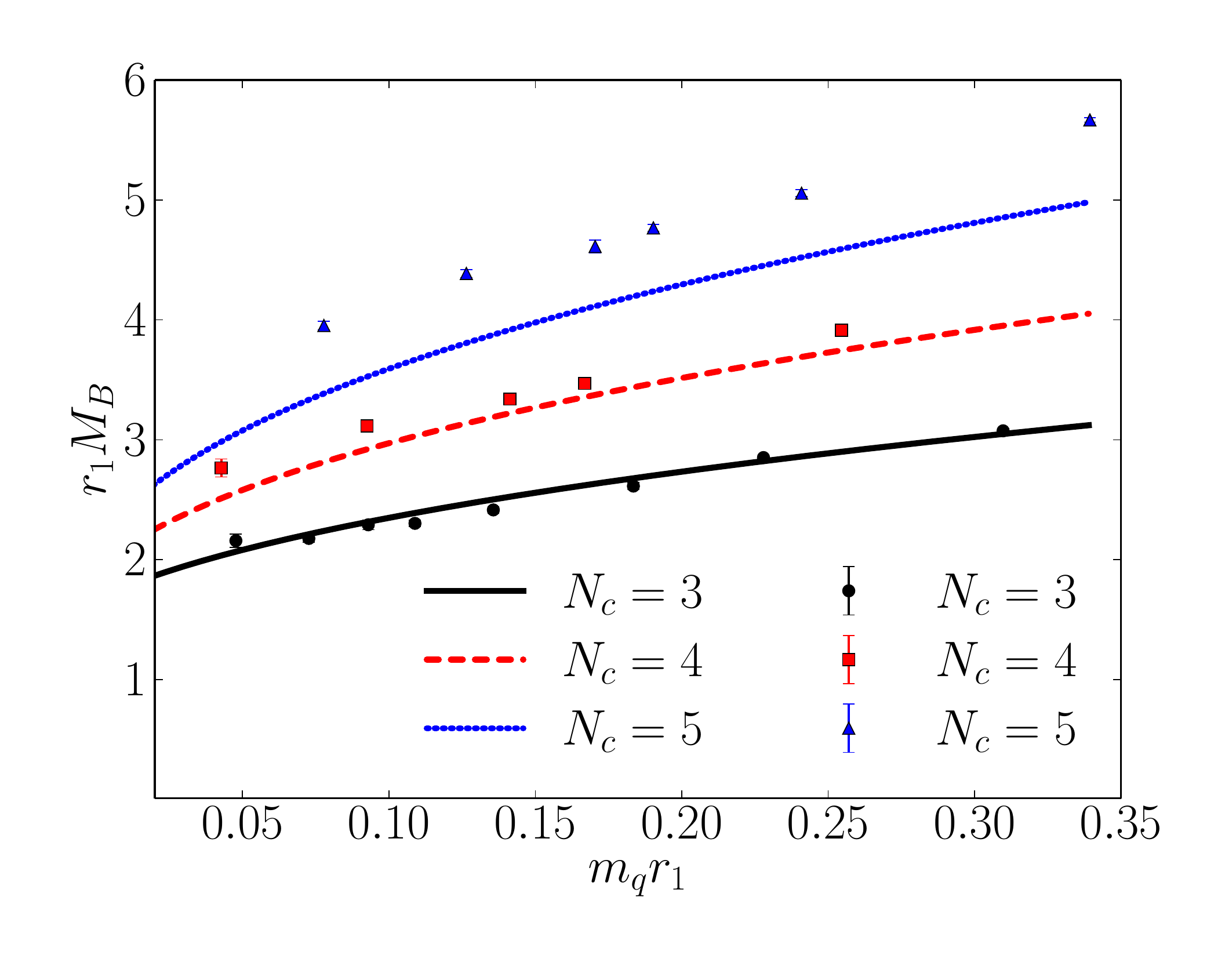}
}
\caption{ The nucleon mass  as a function of the quark mass for 
	$N_c=3$ (left) and its $N_c$ dependence (right).
	The model results are shown by the shaded region which represents the 
	variation of the sigma mass from 950 to 1050 MeV; and by the curves, in this case 
	the sigma mass is taken to be 1000 MeV. 
	The lattice data (symbols) is from Ref.~\cite{DeGrand:2016pur}. 
	The star shows the physical value of the observables.  
}
\label{fig:mn}
\end{figure}

An explicit form for the skyrmion solution is difficult to argue, since one
will have all order in derivatives, as is really the case for the standard
skyrmion solution, and because the $\omega$ meson will play an essential role
in its structure.  Nevertheless, the dependence of the mass of the skyrmion
upon $f_\pi$ can be compared to the 
lattice data, see Fig.~\ref{fig:mn}. The figure demonstrates quite a good
agreement affirming our discussion based
on the skyrmion argument.   

\section{Strength of Interaction}

As described above, the short distance repulsive core is associated with omega
and sigma exchange.
What about the pion range pion tail?  The conventional pion nucleon interaction
of the sigma model,
g $ \overline \psi \tau \cdot \pi \gamma^5 \psi$, has an interaction strength
naively of order $\sqrt{N_c}$ from the coupling.  However, for a
non-relativistic nucleon, this is of order $g/2M_{\rm nucleon} \overline \psi
\gamma^\mu \gamma^5 \psi $ which is of order 1 in powers of $N_c$.  There is
also a potentially dangerous term that arises from the axial current
interaction:
\begin{equation}
 {g^2 \over \Lambda_{\rm QCD}^2} \left\{  \overline \psi_L \chi \gamma \cdot \partial \chi^\dagger \psi_L + \overline \psi _R \chi^\dagger \gamma \cdot \partial \chi \psi_R\right\} .
\end{equation}
Here $g^2$ is of order 1 in powers of $N_c$, and if this interaction is
generated by the exchange of an axial vector meson, then the scale is
$\Lambda_{\rm QCD}$.  For  $f_\pi \sim  O(1)$ in powers of $N_c$ again this
generates an interaction of order 1.  Note that the basic pion nucleon
interaction strength is reduced by a factor of $\sqrt{N_c}$ relative to the naive
counting where $f_\pi \sim \sqrt{N_c}$

This counting means that the one pion exchange isospin dependent interaction is
of order
$1/\Lambda^2_{\rm QCD}R^3$ which at the typical length scale of the nucleon is of
order $1/N_c^{3/2}$.  For isopsin singlet interaction which should be typical of
nuclear matter, two pion exchanges are important, and these are of strength
$1/\Lambda^4_{\rm QCD} R^5 \sim 1/N_c^{5/2}$.  So the picture naturally arises that the nucleon has a strongly repulsive core with a weak long scale interaction generated by pion exchange.  There is no strong long range force.
This is consistent with the phenomenology of nuclear matter.

One can of course also include the effects of massive mesons $M \sim \Lambda_{\rm QCD}$, but such mesons
when convoluted over the large size scale of the nucleon give small effects.

The issue of the binding of nuclear matter is a very subtle one and may be
special to the case of intermediate values of $N_c$ and the details of the
pion, sigma and omega meson masses.  Nevertheless the basic outline of our
description seem reasonable.

Concluding this section, we want to mention that the effective theory 
with the required properties was discussed in  the literature before, see e.g.
~\cite{Meissner:1986ka}. First of all, the vector meson (rho and omega ) masses 
satisfy the KSRF relation $m_V^2 \propto f_\pi^2$. Second, at least for three colors, 
the parameters of the model are consistent with the phenomenologically reasonable hierarchy of the interaction 
ranges; one may expect that the associated hierarchy of the coupling constants between 
matter fields scalar/vector mesons is preserved at larger $N_c$. 

\section{Summary} 
In this article we tried to resolve the issue of the interaction strength  
for the nuclear matter. For this we needed to make assumptions on the 
QCD behaviour in the large $N_c$ limit; namely, in contrast to the conventional 
scaling $N_c^0$, we considered that the 
negative mass term of the associated linear sigma meson Lagrangian is inversely proportional to $N_c$ and thus vanishes 
in the large $N_c$ limit. We showed that this radical assumption does not 
contradict the existent lattice QCD data, which provides results for the quark masses in the range 
where, in our approach,  the conventional scaling 
still holds $m_q\gg \Lambda_{\rm QCD} /N_c^{3/2}$.  With some modest assumptions for the 
values of the scalar-nucleon and vector-nucleon coupling constants, we were able 
to get a weekly attractive nucleon-nucleon potential admitting an interaction energy of 
order the physical scale.

\section{Acknowledgements}
We thank  Robert Pisarski for valuable discussions. 
L. McLerran acknowledges useful conversations with Jean Paul Blaizot and Tory Kojo.  
L. McLerran was  supported under Department of Energy contract number Contract
No. DE-SC0012704 at Brookhaven National Laboratory, and L. McLerran is now
supported under
Department of Energy under grant number DOE grant No. DE-FG02-00ER41132.


\begin{thebibliography} {000}

\bibitem{Witten:1979kh} 
  E.~Witten,
  Nucl.\ Phys.\ B {\bf 160}, 57 (1979).
  doi:10.1016/0550-3213(79)90232-3


\bibitem{DeGrand:2016pur} 
  T.~DeGrand and Y.~Liu,
  Phys.\ Rev.\ D {\bf 94}, no. 3, 034506 (2016)
  doi:10.1103/PhysRevD.94.034506
  [arXiv:1606.01277 [hep-lat]].
  
\bibitem{Bali:2013kia}
  G.~S.~Bali, F.~Bursa, L.~Castagnini, S.~Collins, L.~Del Debbio, B.~Lucini and M.~Panero,
  JHEP {\bf 1306} (2013) 071
  doi:10.1007/JHEP06(2013)071
  [arXiv:1304.4437 [hep-lat]].


  
\bibitem{Adkins:1983ya}
  G.~S.~Adkins, C.~R.~Nappi and E.~Witten,
  Nucl.\ Phys.\ B {\bf 228} (1983) 552.
  doi:10.1016/0550-3213(83)90559-X
  
\bibitem{Klebanov:1985qi}
  I.~R.~Klebanov,
  Nucl.\ Phys.\ B {\bf 262} (1985) 133.
  doi:10.1016/0550-3213(85)90068-9
  
\bibitem{Hidaka:2010ph} 
  Y.~Hidaka, T.~Kojo, L.~McLerran and R.~D.~Pisarski,
  Nucl.\ Phys.\ A {\bf 852}, 155 (2011)
  doi:10.1016/j.nuclphysa.2011.01.008
  [arXiv:1004.2261 [hep-ph]].


\bibitem{Adam:2010ds}
  C.~Adam, J.~Sanchez-Guillen and A.~Wereszczynski,
  Phys.\ Rev.\ D {\bf 82} (2010) 085015
  doi:10.1103/PhysRevD.82.085015
  [arXiv:1007.1567 [hep-th]].






  
  
  
  
\bibitem{Wess:1971yu} 
  J.~Wess and B.~Zumino,
  Phys.\ Lett.\  {\bf 37B}, 95 (1971).
  doi:10.1016/0370-2693(71)90582-X

\bibitem{Witten:1983tw} 
  E.~Witten,
  Nucl.\ Phys.\ B {\bf 223}, 422 (1983).
  doi:10.1016/0550-3213(83)90063-9
  
\bibitem{Bonenfant:2010ab}
  E.~Bonenfant and L.~Marleau,
  Phys.\ Rev.\ D {\bf 82} (2010) 054023
  doi:10.1103/PhysRevD.82.054023
  [arXiv:1007.1396 [hep-ph]].


\bibitem{Adam:2011zz}
  C.~Adam, J.~Sanchez-Guillen and A.~Wereszczynski,
  AIP Conf.\ Proc.\  {\bf 1343} (2011) 598.
  doi:10.1063/1.3575106
  
\bibitem{tHooft:1974pnl} 
  G.~'t Hooft,
  Nucl.\ Phys.\ B {\bf 75}, 461 (1974).
  doi:10.1016/0550-3213(74)90088-1

  
\bibitem{Coleman:1980mx}
  S.~R.~Coleman and E.~Witten,
  Phys.\ Rev.\ Lett.\  {\bf 45} (1980) 100.
  doi:10.1103/PhysRevLett.45.100
  
\bibitem{Meissner:1986ka} 
  U.~G.~Meissner, N.~Kaiser, A.~Wirzba and W.~Weise,
  Phys.\ Rev.\ Lett.\  {\bf 57}, 1676 (1986).
  doi:10.1103/PhysRevLett.57.1676

\end{thebibliography}
\end{document}